\documentclass[10pt,a4paper,superscriptaddress,aps,prd,nofootinbib,reprint]{revtex4-2}

\usepackage[utf8]{inputenc}
\usepackage{amsmath,amsfonts,amssymb}
\usepackage{graphicx}
\usepackage[caption=false]{subfig} 
\usepackage[breaklinks,colorlinks]{hyperref}
\usepackage[capitalise]{cleveref} 
\usepackage{microtype}
\usepackage{xcolor}
\usepackage{academicons}
\usepackage{fontawesome5}
\usepackage{xspace}

\definecolor{orcidlogocol}{rgb}{0.65, 0.807, 0.223}
\newcommand{\orcid}[1]{$\,$\href{https://orcid.org/#1}{\textcolor{orcidlogocol}{\faOrcid}}}

\def\Planck{\textit{Planck}\xspace}

\hypersetup{
  urlcolor=blue,
  citecolor=blue,
  linkcolor=blue
}

\begin{document}

\title{How isotropic is dark energy?}
 
\author{Richard A. Battye}
\email[]{richard.battye@manchester.ac.uk}
\affiliation{%
Jodrell Bank Centre for Astrophysics, Department of Physics and Astronomy, University of Manchester, Manchester, M13 9PL, U.K.
}

\author{Adam Moss\orcid{0000-0002-7245-7670}}
\email[]{adam.moss@nottingham.ac.uk}
\affiliation{%
School of Physics and Astronomy, University of Nottingham, Nottingham, NG7 2RD, U.K.
}

\label{firstpage}

\date{\today}

\begin{abstract}
Tensions in late-time expansion data have renewed interest in models beyond $\Lambda$CDM. We ask: \emph{how isotropic must dark energy be?} 
Working in Bianchi~I, we allow time-dependent anisotropic stress and introduce a parameterisation that enforces a vanishing line-of-sight integral of the shear, thereby satisfying the CMB ISW quadrupole bound by construction. 
Using Pantheon+SH0ES SNe together with DESI BAO distances, single-bin (constant) and five-bin anisotropic models improve the fit over $w$CDM by $\Delta(-2\ln\mathcal{L}_{\rm iso})=14.8$ and $26.6$ respectively, but both violate the quadrupole constraint. 
In contrast, a five-bin constrained model achieves $\Delta(-2\ln\mathcal{L}_{\rm iso})=15.4$ while remaining compatible with the quadrupole limit. 
The fit improvement arises from two sources: capturing directional structure in the Pantheon+ SNe data, and partially alleviating the tension between the SH0ES $H_0$ value and DESI BAO distances.
\end{abstract}
      

\maketitle

\section{Introduction}

The origin of the dark energy that drives the observed acceleration~\cite{SupernovaSearchTeam:1998fmf,SupernovaCosmologyProject:1998vns} is one of the great mysteries of modern cosmology~\cite{Copeland2006}. 
Within the framework of the homogeneous and isotropic Friedmann-Robertson-Walker (FRW) model, acceleration requires the density, $\rho_{\rm de}$, and pressure, $P_{\rm de}$, to have an equation of state parameter $w=P_{\rm de}/\rho_{\rm de}<-\textstyle{\frac{1}{3}}$ which violates the Strong Energy Condition (SEC). 
For many years the consensus has been that this dark energy comes from a cosmological constant, $\Lambda$, with $w=-1$.

Recent observations from the Dark Energy Spectroscopic Instrument (DESI)~\cite{DESI:2025zgx,DESI:2025fii} when combined with those from the \Planck Satellite~\cite{Planck:2015fie,Planck:2018vyg} and of Type Ia Supernovae (SNe)~\cite{Scolnic:2021amr,DES:2024jxu,Rubin:2023jdq} have been used to claim evidence that $w$ is not only time-varying, but that $w<-1$ is required at some point during the evolution of the Universe.
This is a significant challenge to the orthodoxy of the $\Lambda$CDM model, even though combining so many probes requires care since it can amplify any systematic uncertainties. 

Given these tensions in the standard model, it is worth exploring alternative frameworks. Anisotropic dark energy models represent one such possibility, allowing the pressure to vary with direction. 
In the Bianchi I cosmological framework, dark energy can exhibit different pressures along different spatial axes, characterized by an anisotropic stress tensor. 
Previous observational constraints on cosmic anisotropy have been derived from diverse probes: CMB temperature and polarization anisotropies limit large-scale shear to $\sigma_0/H_0 \lesssim 10^{-10}$~\cite{Saadeh:2016sak}, while Type Ia SNe analyses have found evidence for preferred directions at various significance levels~\cite{Appleby:2015gla,Colin:2019opn}, with recent Pantheon+ studies reporting dipole~\cite{Sorrenti:2022zat,Sah:2024csa} and quadrupole~\cite{Cowell:2022ehf} signatures. 
However, direct fits of anisotropic Bianchi models to SNe data yield much larger shear parameters~\cite{Verma:2024lex}, creating tension with early-universe constraints.
The challenge for anisotropic dark energy models is to produce observable late-time signatures while remaining consistent with the stringent constraints from the CMB~\cite{Appleby:2009za}.

Several dark-energy and modified gravity models admit spatially homogeneous but anisotropic backgrounds, providing a microphysical rationale for a time-dependent anisotropic equation of state $\Delta w(a)$. 
One example is ``solid'' (elastic) dark energy\footnote{Early work on solid dark energy~\cite{Battye:2006mb,Battye:2009ze} considered an isotropic background but anisotropic perturbations, due to a direction dependent sound speed.}, built from a triad of scalar fields arranged so that homogeneity is preserved even though rotational invariance is broken, which can yield anisotropic accelerated attractors~\cite{Motoa-Manzano:2020mwe}. 
Other possibilities include spatial vector-field dark energy, where a spacelike background vector picks out a preferred axis and the resulting Bianchi~I system exhibits scaling and accelerating solutions in dynamical-systems analyses~\cite{Koivisto:2008xf}. 
Higher-form fields provide another route: interacting two-form models have been studied directly in a Bianchi~I background and can admit late-time de~Sitter-like attractors with non-trivial anisotropy~\cite{Orjuela-Quintana:2022jrg}. 
Finally, preferred-frame theories such as Einstein--\ae{}ther constructions also support homogeneous anisotropic cosmologies, including Bianchi~I solutions~\cite{Alhulaimi:2013sha}. 

In this paper we explore time-dependent anisotropic dark energy within the Bianchi I framework. 
We develop a novel parameterization that circumvents the stringent CMB constraints through a constrained basis function approach that automatically ensures the integrated shear vanishes over cosmic time. 
We then apply this framework to fit the complete Pantheon+SH0ES SNe dataset, constraining both the amplitude and temporal evolution of anisotropic dark energy. 

\section{Bianchi I universe}

The Bianchi classification can be used to describe homogeneous, anisotropic, expanding spacetimes~\cite{Ellis:1968vb}. For the purposes of our discussion we will use the Bianchi I model which can be thought of as having non-isotropic pressure and different scale factors in each of the 3 coordinate directions. It can be formulated using the flat metric~\cite{Pereira:2007yy}
\begin{equation}
    ds^2 = a(\eta)^2\left(-d\eta^2 + \gamma_{ij}(\eta)\,dx^i dx^j\right)\,.
\end{equation}
The Friedmann equation is
\begin{equation}
    3{\cal H}^2=8\pi Ga^2\rho_{\rm tot}+\textstyle{1\over 2}\sigma^2\,,
\end{equation}
where $\sigma_{ij}=\textstyle{1\over 2}\gamma_{ij}^\prime$, $\sigma^2=\sigma_i^j\sigma_j^i$ and ${\cal H}=a^\prime/a$ where $^\prime$ denotes derivative with respect to $\eta$. We will consider a Universe where the total density is $\rho_{\rm tot}=\rho_{\rm m}+\rho_{\rm de}$ where $\rho_{\rm m}\propto a^{-3}$ is the matter density. 

The pressure tensor for the dark energy will be assumed to be anisotropic and of the form
\begin{equation}
    P_i^j = \rho_{\rm de}\left[w_{\rm de}\delta_i^j+\Delta w_i^j\right]\,,
\end{equation}
where $w_{\rm de}$ is a constant, but the traceless anisotropic equation of state parameter $\Delta w_i^j\equiv\Delta w_i^j(\eta)$ is time-dependent. 
The dark energy density $\rho_{\rm de}$ and shear $\sigma_i^j$ satisfy~\cite{Appleby:2009za}
\begin{eqnarray}
    \rho_{\rm de}^{\prime}&=&-3{\cal H}(1+w_{\rm de})\rho_{\rm de}-\sigma_i^j\Delta w_j^i\rho_{\rm de}\,,\cr
    \sigma_i^{j\,\prime}&=&-2{\cal H}\sigma_i^j+8\pi Ga^2\Delta w_i^j\rho_{\rm de}\,.
\end{eqnarray}

Without loss of generality we choose the principal-axis frame, aligning the symmetry axis with $\hat z$ (i.e. $\hat\lambda\parallel\hat z$), so that
$\Delta w_i{}^{j}=\mathrm{diag}(\Delta w_1,\Delta w_2,-[\Delta w_1+\Delta w_2])$. Allowing time variation provides significant freedom in choosing these components, which we exploit in our parameterization. 

\section{Integrated Sachs-Wolfe (ISW) effect}

Anisotropic dark energy models generate an ISW effect in the CMB through the integrated shear along photon paths. The temperature anisotropy is given by
\begin{equation}
    {\Delta T\over T}=-\int_{\eta_{\rm rec}}^{\eta_0}\sigma_{ij}(\eta)n^{i}n^{j}d\eta\,,
\end{equation}
where $n^i$ is the line-of-sight direction, $\eta_0$ is the present conformal time, and $\eta_{\rm rec}$ is the conformal time at recombination. This generates a CMB temperature quadrupole with coefficients $a_{2m}^{\rm aniso}$. 
Treating these as a 5-vector, the inflationary and anisotropic contributions add with an arbitrary relative orientation. Since the observed quadrupole is lower than the $\Lambda$CDM prediction, cancellation between these contributions is possible, implying the envelope
\begin{equation}
\bigl(\sqrt{D_2^{\rm inf}}-\sqrt{D_2^{\rm obs}}\bigr)^2
\;\le\;
D_2^{\rm aniso}
\;\le\;
\bigl(\sqrt{D_2^{\rm inf}}+\sqrt{D_2^{\rm obs}}\bigr)^2,
\end{equation}
where superscripts “obs/inf/aniso” denote observed, inflationary, and anisotropic values, and $D_\ell \equiv \ell(\ell+1) C_\ell / (2\pi)$.
Using $D_2^{\rm obs}\approx 226~\mu\mathrm{K}^2$ (Planck DR3~\cite{Planck:2019nip}) and $D_2^{\rm inf}\approx 1000~\mu\mathrm{K}^2$~($\Lambda$CDM), this yields an approximate range 
\begin{equation}
D_2^{\rm aniso}\in\bigl[\,275\,,\,2177\,\bigr]~\mu\mathrm{K}^2.
\end{equation}
This envelope places a restrictive upper bound on anisotropic contributions, yet it has been neglected in recent SNe-based analyses~\cite{Verma:2024lex}.

\section{Parameterization}

We choose the axisymmetric case $\Delta w_1 = \Delta w_2 = \Delta w(a)$ and parameterize $\Delta w(a)$ using a piecewise constant representation (bins) over the interval $a \in [a_*, a_0]$. We divide the interval into $N$ bins with contiguous top-hat functions $B_k(a)$
\begin{equation}
    \label{eq:dw_bins}
    \Delta w(a) = \sum_{k=0}^{N-1} c_k B_k(a)\,,
\end{equation}
where $c_k$ represents the constant value of $\Delta w(a)$ in the $k$-th bin. We will choose $a_* = 1/4$ (corresponding to $z_*=3$) as this is the region most relevant for dark energy. 

In the axisymmetric case, the shear in its principal frame is
$\sigma_i{}^{j}=s(a)\,\mathrm{diag}(1,1,-2)$, yielding a CMB pattern with only $a_{20}^{\rm aniso}$ non-zero. Defining
\begin{equation}
  Q_T(\eta_0)\;\equiv\;-\!\int_{\eta_*}^{\eta_0} s(\eta)\,d\eta\,,
\end{equation}
we obtain
\begin{equation}
\frac{\Delta T}{T}(\hat n)= -2 Q_T \,P_2(\cos\zeta),\qquad
C_2^{\rm aniso}=\frac{16\pi}{25}\,Q_T^2\,,
\end{equation}
where $\zeta = \cos^{-1}(\hat{\lambda} \cdot \hat{n})$ is the angle between the line-of-sight direction $\hat{n}$ and $\hat{\lambda}$, and $C_2^{\rm aniso}$ is in $(\Delta T/T)^2$ units. 
Taking the most conservative limit above, this implies $C_2^{\rm aniso} \lesssim 3.1 \times 10^{-10}$. For constant $\Delta w$, this imposes a stringent constraint of $\Delta w \lesssim 10^{-4}$~\cite{Appleby:2009za}. 
However, this can be circumvented with time-varying anisotropy if the $c_k$ coefficients have different signs, leading to cancellation along the line-of-sight (i.e.\ $Q_T \approx 0$). This inevitably involves some tuning of the coefficients, so to ensure our parameterization satisfies the quadrupole constraint without having an extremely low MCMC acceptance rate, we develop a constrained basis function approach that enforces $Q_T = 0$ in Appendix~\ref{sec:cons}. This results in a set of amplitude coefficients $\boldsymbol{\alpha}\in\mathbb{R}^{N-1}$ for the $N$ bins.

\section{Distance measures}

In anisotropic spacetimes, the luminosity distance becomes direction-dependent. The ellipticity parameter $e^2$ quantifies the deviation from isotropy and evolves in the axisymmetric case according to
\begin{equation}
    \left( e^2 \right)^{\prime} = 6 s ( 1 - e^2)\,.
\end{equation}
The luminosity distance in a Bianchi I universe is given by~\cite{Verma:2024lex}
\begin{equation}
    d_L(\hat{n}, z) = (1 + z) \int_{0}^z \frac{dz'}{H(z')} \frac{(1 - e^2)^{1/6}}{(1 - e^2\cos^2\zeta)^{1/2}}\,,
   \label{eq:lumdist}
\end{equation}
where $H$ is the physical Hubble rate.

For BAO distances, we compare to DESI's measurements using the standard FRW
   definitions in terms of the Hubble rate and the drag-epoch sound
   horizon \(r_{\rm drag}\). For a spatially flat background,
\begin{equation}
  D_M(z)=\int_{0}^{z}\frac{dz'}{H(z')}\,,\qquad
  D_H(z)=\frac{1}{H(z)}\,,
  \label{eq:DM_DH_DV}
\end{equation}
where $D_V(z)=\big[z\,D_H(z)\,D_M^2(z)\big]^{1/3}$. The observables used in the likelihood are the dimensionless ratios $D_M(z) / r_{\rm drag}$, $D_H(z) / r_{\rm drag}$, and $D_V(z) / r_{\rm drag}$.
   
We use isotropic (compressed) BAO distances and defer a full
directional treatment to future work\footnote{This could be done, for example, by treating the SGC and NGC separately.}. This
constrains the average background \(H(z)\), so includes changes
from (i) modified scaling of \(\rho_{\rm de}\) and (ii) the shear energy
density \(\rho_\sigma\!\propto\!\sigma^2/a^2\) entering \(H^2\). 
This is a conservative choice for constraining anisotropy: a dedicated directional BAO likelihood would generically add sensitivity to anisotropic expansion (and could therefore tighten constraints), whereas the isotropic compression largely washes out such information.

\begin{figure}[t!]
  \centering
  \includegraphics[width=0.45\textwidth]{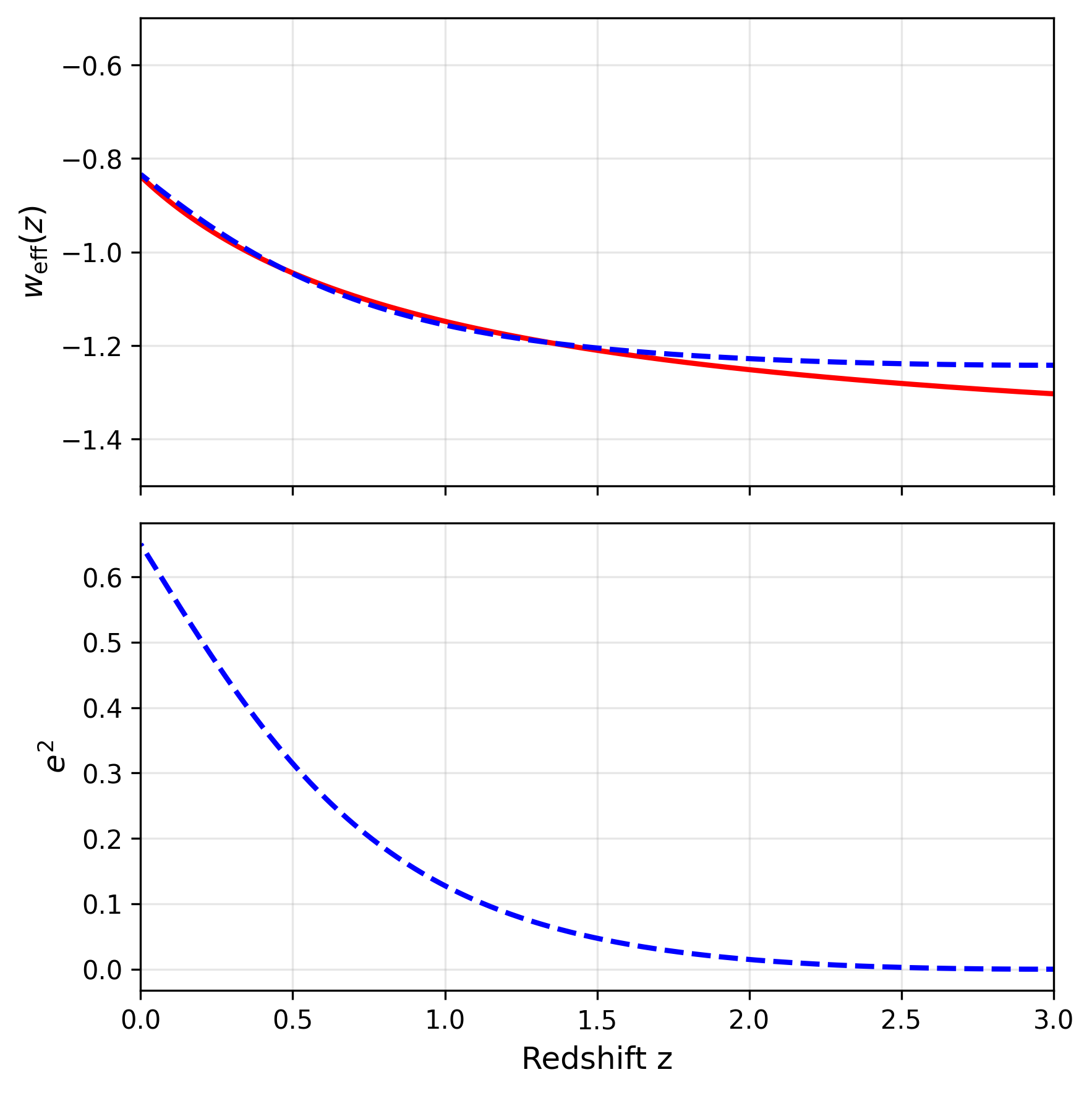}
  \caption{Effective equation of state $w_{\rm eff}(a)$ as a function of scale factor. Comparison of a dynamical dark energy model (solid red) with $w(a)=w_0+w_a(1-a)$ and the single-bin anisotropic model (dashed blue).}
  \label{fig:cpl_mimic}
  \end{figure}

Motivated by DESI results, we ask whether late-time anisotropy can mimic the background expansion favored by dynamical dark energy, with an equation of state of the form $w(a)=w_0+w_a(1-a)$. 
In our framework, one can define an effective dark-energy density---including the shear---and equation of state that encode the average background
\begin{eqnarray} \label{eq:rho_de_eff}
  \rho_{\rm de}^{\rm (eff)} &=& \rho_{\rm de}+{\sigma^2\over 16\pi Ga^2}\,, \\
w_{\mathrm{eff}}(a) &=&
   -1-\frac{a}{3\,\rho_{\mathrm{de}}^{\mathrm{(eff)}}}
       \frac{d\rho_{\mathrm{de}}^{\mathrm{(eff)}}}{da}\,.
\end{eqnarray}
One can show that $w_{\rm eff} \geq w_{\rm de}$ for $w_{\rm de} \leq 1$ (see Appendix~\ref{sec:weff}), so the shear contribution always stiffens the effective equation of state.
For illustration, we take a representative fit\footnote{These are marginalized posterior values from the DESI analysis for the data combination DESI+CMB+Pantheon+.} with $(w_0,w_a)=(-0.838,-0.62)$ and ask whether a \emph{single-bin} anisotropic model can reproduce the same $H(z)$.
With axisymmetry and constant parameters $(w_{\rm de},\Delta w) = (-1.242,\,0.706)$, the resulting $w_{\rm eff}(a)$ closely follows the dynamical dark energy over $z\in[0,3]$, as shown in Fig.~\ref{fig:cpl_mimic}.
However, this model predicts a quadrupole power $C_2^{\rm aniso} = 6.2\times10^{-2}$, together with a large ellipticity $e^2$ at low redshift, which may be in conflict with SNe data and other local measurements. 
The binned approach offers greater scope to approximate dynamical dark energy while remaining compatible with the CMB quadrupole constraint.

\section{Methods and Data}

We analyze the complete Pantheon+SH0ES compilation~\cite{Scolnic:2021amr}\footnote{\url{https://github.com/PantheonPlusSH0ES/DataRelease}} containing 1701 Type Ia SNe light curves spanning redshifts $0.001 < z < 2.26$. 
This dataset represents the largest and most homogeneous SNe Ia sample to date, incorporating significant improvements including enhanced low-redshift coverage, refined calibration procedures, and updated systematic uncertainty treatments. 
The sample includes SH0ES Cepheid distance calibrators for 42 SNe Ia in nearby galaxies that anchor the distance scale.

We extract the raw SALT2 light curve parameters: peak B-band apparent magnitude $m_B$, stretch parameter $x_1$, and color parameter $c$, along with their uncertainties and host galaxy stellar masses. 
Sky positions are converted from equatorial to Galactic coordinates using HEALPix~\cite{Gorski:2004by} to compute directional cosines for anisotropy analysis.

The standardized apparent B-band magnitude incorporates empirical corrections: $m_{B,\rm corr} = m_B + \alpha x_1 - \beta c + \gamma G_{\rm host}$, where $\alpha$, $\beta$, and $\gamma$ are global nuisance parameters accounting for the Phillips relation (stretch-luminosity), dust extinction (color-luminosity), and host galaxy stellar mass bias, respectively. 
The mass step function is $G_{\rm host} = +0.5$ for $M_{\rm host} > 10^{10} M_\odot$ and $-0.5$ otherwise.

Our theoretical framework treats Cepheid calibrators and cosmological SNe differently. For the 42 Cepheid host galaxies, we use the provided geometric distance moduli $\mu_{\rm Ceph}$ directly, bypassing cosmological distance calculations. 
For cosmological SNe, we compute the theoretical distance modulus as $\mu_{\rm th} = 5\log_{10}[(1+z_{\rm CMB})(1+z_{\rm hel})D_A] + 25$, where $D_A$ is the angular diameter distance derived from the anisotropic luminosity distance via $D_A = d_L(\hat{n},z)/(1+z_{\rm CMB})^2$. 
This formulation accounts for both the CMB dipole correction ($z_{\rm CMB}$) and heliocentric motion ($z_{\rm hel}$), following the Pantheon+ analysis prescription, but adapted for directionally-dependent distances in the Bianchi framework.

We incorporate baryon acoustic oscillation (BAO) measurements from the Dark Energy Spectroscopic Instrument (DESI) Year 2 data release~\cite{DESI:2025zgx}. The dataset includes 13 BAO measurements spanning redshifts $z = 0.295$ to $z = 2.33$, comprising: one isotropic distance measurement $D_V/r_{\rm drag}$ at $z = 0.295$ from the bright galaxy sample (BGS), and twelve measurements of the transverse ($D_M/r_{\rm drag}$) and radial ($D_H/r_{\rm drag}$) BAO scales from luminous red galaxies (LRGs) at $z = 0.51, 0.706, 0.934$, emission line galaxies (ELGs) at $z = 1.321, 1.484$, and quasars (QSOs) at $z = 2.33$. 
The BAO likelihood uses the full $13\times13$ covariance matrix accounting for statistical and systematic uncertainties. We numerically marginalize over the sound horizon $r_{\rm drag}$ using the Planck prior $r_{\rm drag} = (147.09 \pm 0.26)\,{\rm Mpc}$.

We adopt flat priors reflecting physical constraints and previous measurements: $\Omega_{\rm de} \in [0.1,0.9]$ (dark energy density relative to the critical density today), $w_{\rm de} \in [-2.0,-0.33]$ (equation of state parameter), $H_0 \in [50,90]\,{\rm km\,s^{-1}\,Mpc^{-1}}$ (Hubble constant), and $M \in [-19.5,-18.5]$ (absolute magnitude). 
The SALT2 nuisance parameters use $5\sigma$ ranges centered on Pantheon+ values: $\alpha \in [0.10,0.18]$, $\beta \in [2.0,2.9]$, and $\gamma \in [-0.05,0.12]$. For anisotropic models, the preferred axis direction uses $\theta_{\rm axis} \in [0,\pi/2]$ (polar angle, restricted to one hemisphere due to axisymmetry) and $\phi_{\rm axis} \in [0,2\pi]$ (azimuthal angle) in Galactic coordinates. 
SVD mode coefficients are constrained to $\alpha_j \in [-1,1]$ for $j=1,\ldots,N-1$.

The likelihood function is $\mathcal{L} \propto \exp(-\chi^2/2)$ where $\chi^2 = \chi^2_{\rm SN} + \chi^2_{\rm BAO}$. The SNe contribution is $\chi^2_{\rm SN} = \Delta\boldsymbol{\mu}^T \mathbf{C}_{\rm SN}^{-1} \Delta\boldsymbol{\mu}$ with residual vector components $\Delta\mu_i = m_{B,\rm corr}^i - M - \mu_{\rm th}^i$ for cosmological SNe and $\Delta\mu_i = m_{B,\rm corr}^i - M - \mu_{\rm Ceph}^i$ for Cepheid hosts. 
The SN covariance matrix $\mathbf{C}_{\rm SN} = \mathbf{C}_{\rm stat+sys}^{\rm SN} + \mathbf{C}_{\rm stat+sys}^{\rm Ceph}$ incorporates both statistical uncertainties from light curve fitting and systematic uncertainties including calibration, host galaxy corrections, and dust extinction models. The BAO contribution is $\chi^2_{\rm BAO} = (\boldsymbol{d}_{\rm obs} - \boldsymbol{d}_{\rm th})^T \mathbf{C}_{\rm BAO}^{-1} (\boldsymbol{d}_{\rm obs} - \boldsymbol{d}_{\rm th})$, where $\boldsymbol{d}_{\rm obs}$ and $\boldsymbol{d}_{\rm th}$ are the observed and theoretical BAO distance measurements, respectively, and $\mathbf{C}_{\rm BAO}$ is the $13\times13$ BAO covariance matrix.

We sample the posterior distribution using the affine-invariant ensemble sampler \texttt{emcee}~\cite{Foreman-Mackey:2012ig} with the number of walkers equal to twice the number of model parameters.
Because we employ an affine-invariant ensemble sampler, we assess convergence using the integrated autocorrelation time, $\tau$, and the effective sample size (ESS) rather than Gelman–Rubin across walkers. We estimate $\tau$ from the unthinned chains and continue sampling until the post-burn-in length per walker satisfies $N_{\rm step}\!\ge\!50\,\tau_{\max}$; we also require a minimum ESS per parameter (typically $\gtrsim 10^3$) before reporting posterior summaries.
As a robustness check against survey geometry or footprint-induced effects, we run permutation-based null tests that randomly shuffle sky positions within survey subsets while keeping all other observables and covariances fixed; see Appendix~\ref{sec:null}.

\section{Results}

\begin{figure}[t!]
    \centering
    \includegraphics[width=0.45\textwidth]{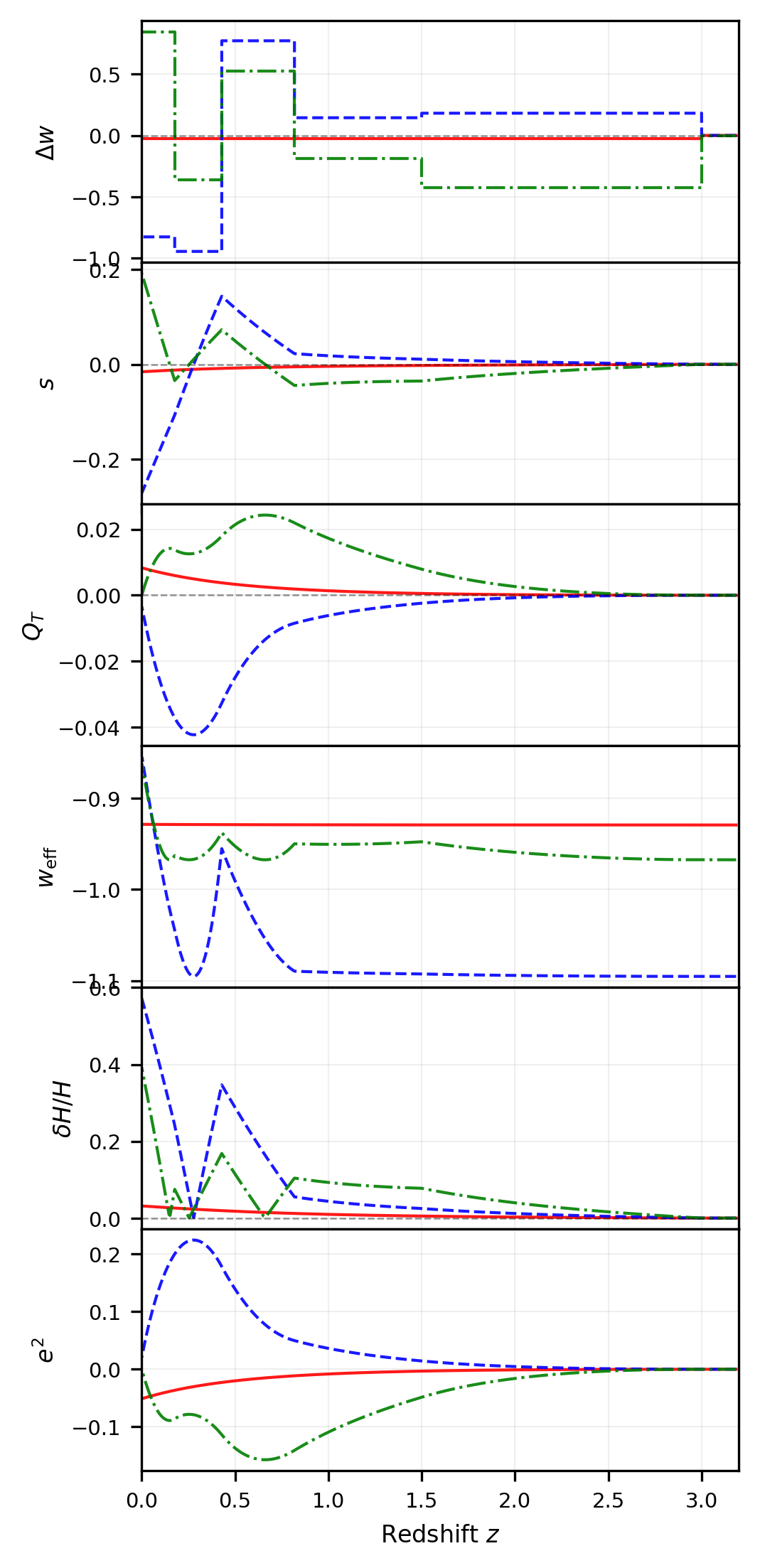}
    \caption{Best-fit evolution of $w_{\rm eff}(a)$, shear $s(a)$, and ellipticity $e^2(a)$ for the constant anisotropic model (solid red), 5-bin unconstrained model (dashed blue), and 5-bin SVD model (dot-dashed green).}
    \label{fig:evolution}
    \end{figure}

We present constraints from fitting four cosmological models to the Pantheon+SH0ES and DESI BAO datasets: standard isotropic $w$CDM, single-bin (constant) anisotropic dark energy, five-bin unconstrained anisotropy, and five-bin SVD parameterization with continuation method. 
Table~\ref{tab:results} summarizes the parameter constraints and goodness-of-fit statistics for each model, and best-fit evolutions of key background/anisotropy quantities are presented in Fig.~\ref{fig:evolution}.
The derived quantity $\langle\delta H/H\rangle_{z<0.2}$ represents the RMS directional variation of the Hubble rate at low redshift, defined as $\langle\delta H/H\rangle = \sqrt{4/5}\,s/\mathcal{H}$, which quantifies the observable anisotropy in local expansion.

\begin{table*}[t!]
  \centering
  \caption{Parameter constraints from fitting the Pantheon+SH0ES and DESI BAO datasets to different cosmological models. We show 68\% confidence intervals for all parameters. 
  The $w$CDM model allows constant dark energy equation of state but no anisotropy. The anisotropic models use 1-bin and 5-bin parameterizations, with the SVD model enforcing the CMB constraint through constrained basis functions. 
  The total $\Delta(-2\ln\mathcal{L}_{\rm iso})$ is slightly different from the combined Pantheon+SH0ES + DESI $\chi^2$ values due to the Gaussian prior on $r_{\rm drag}$.}
  \label{tab:results}
  \renewcommand{\arraystretch}{1.3}
  \begin{tabular*}{\textwidth}{@{\extracolsep{\fill}}lcccc@{}}
  \hline\hline
  Parameter & Isotropic $w$CDM & 1-bin Unconstrained & 5-bin Unconstrained & 5-bin SVD \\
  \hline
  \multicolumn{5}{l}{\it Cosmological Parameters} \\
  $\Omega_{\rm de}$ & $0.711_{-0.008}^{+0.008}$ & $0.711_{-0.008}^{+0.008}$ & $0.708_{-0.008}^{+0.008}$ & $0.710_{-0.007}^{+0.010}$ \\
  $w_{\rm de}$ & $-0.925_{-0.037}^{+0.037}$ & $-0.927_{-0.038}^{+0.037}$ & $-1.008_{-0.068}^{+0.056}$ & $-0.974_{-0.056}^{+0.048}$ \\
  $H_0\,[{\rm km\,s^{-1}\,Mpc^{-1}}]$ & $68.70_{-0.54}^{+0.55}$ & $68.70_{-0.52}^{+0.52}$ & $71.28_{-1.46}^{+1.74}$ & $69.88_{-0.85}^{+1.40}$ \\
  \hline
  \multicolumn{5}{l}{\it Anisotropy Parameters} \\
  $c_0$, $\alpha_0$ & --- & $-0.022_{-0.010}^{+0.051}$ & $0.403_{-0.449}^{+0.398}$ & $-0.016_{-0.525}^{+0.388}$ \\
  $c_1$, $\alpha_1$ & --- & --- & $0.076_{-0.476}^{+0.449}$ & $-0.013_{-0.350}^{+0.620}$ \\
  $c_2$, $\alpha_2$ & --- & --- & $0.130_{-0.571}^{+0.523}$ & $0.198_{-0.362}^{+0.323}$ \\
  $c_3$, $\alpha_3$ & --- & --- & $-0.367_{-0.406}^{+0.548}$ & $0.365_{-0.455}^{+0.298}$ \\
  $c_4$ & --- & --- & $-0.639_{-0.266}^{+0.570}$ & --- \\
  $\theta_{\rm axis}$ [rad] & --- & $0.518_{-0.264}^{+0.839}$ & $1.427_{-0.230}^{+0.069}$ & $0.293_{-0.217}^{+0.408}$ \\
  $\phi_{\rm axis}$ [rad] & --- & $0.986_{-0.497}^{+3.828}$ & $4.971_{-3.048}^{+0.110}$ & $5.851_{-4.793}^{+0.342}$ \\
  \hline
  \multicolumn{5}{l}{\it Derived Quantities} \\
  $\log_{10}(C_2^{\rm aniso})$ & --- & $-3.86_{-0.33}^{+0.23}$ & $-4.29_{-1.04}^{+0.59}$ & $-10.05_{-0.84}^{+0.48}$ \\
  $\langle\delta H/H\rangle_{z<0.2}$ & --- & $0.013_{-0.004}^{+0.004}$ & $0.10_{-0.04}^{+0.03}$ & $0.05_{-0.02}^{+0.02}$ \\
  \hline
  \multicolumn{5}{l}{\it Goodness of Fit} \\
  $\chi^2_{\rm Pantheon+SH0ES}$ & 1678.93 & 1663.74 & 1650.88 & 1664.78  \\
  $\chi^2_{\rm DESI}$ & 11.01 & 11.65 & 12.93 & 10.11 \\
  $\Delta(-2\ln\mathcal{L}_{\rm iso})$ & 0 & 14.75 & 26.64 & 15.37 \\
  \hline\hline
  \end{tabular*}
  \end{table*}

    \begin{figure}[t!]
      \centering
      \includegraphics[width=0.45\textwidth]{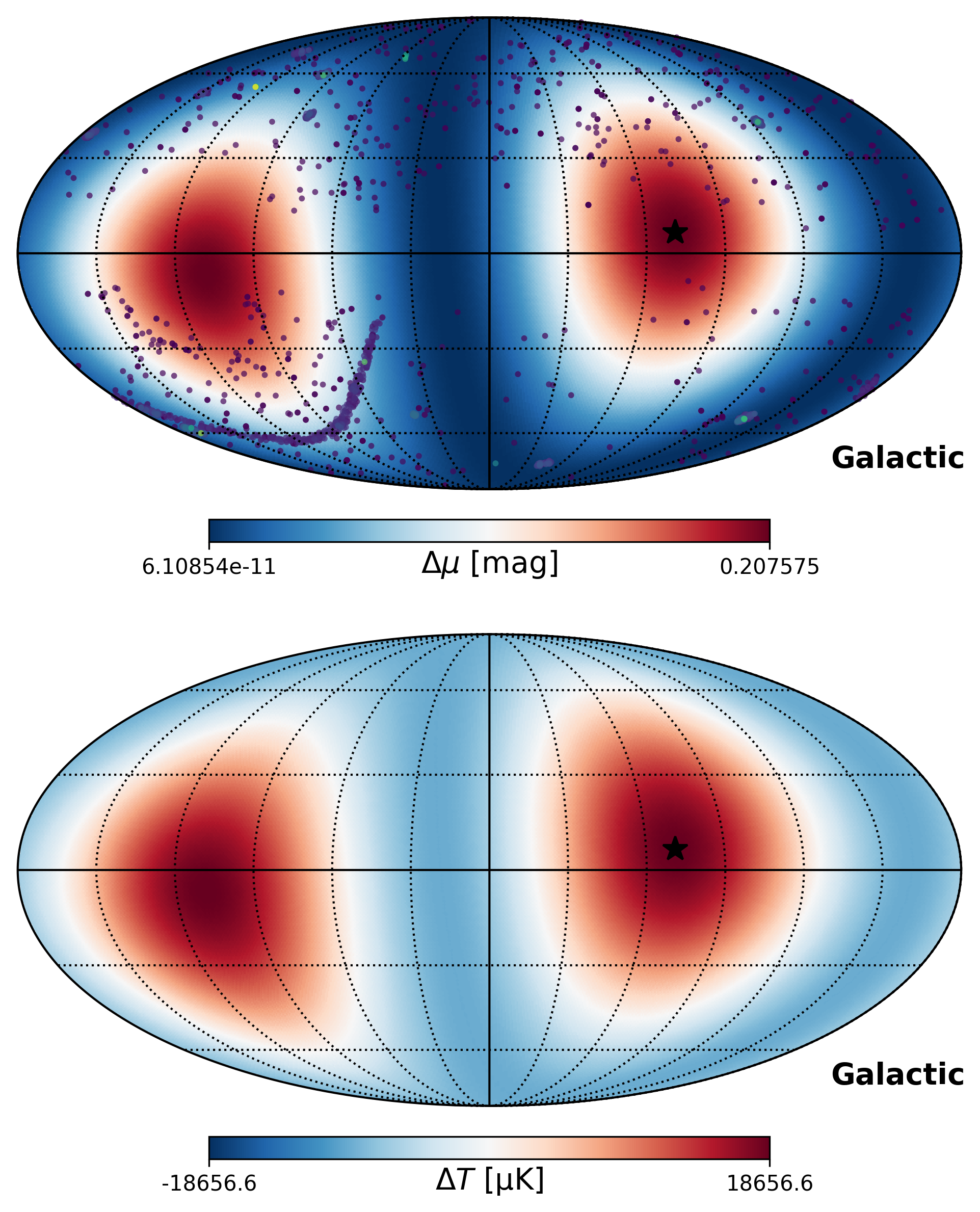}
      \caption{Anisotropy sky maps for the 5-bin unconstrained model. \textit{Top panel}: Directional variation in distance modulus at redshift $z=0.5$, shown in Galactic coordinates with the preferred anisotropy axis marked by a black star. White dots indicate Pantheon+SH0ES SNe positions. \textit{Bottom panel}: Predicted CMB temperature quadrupole pattern. The best-fit anisotropy axis points toward Galactic coordinates $(l,b) = (288.8^{\circ}, 6.5^{\circ})$, consistent with the preferred direction found in~\cite{Verma:2024lex}.}
      \label{fig:anisotropy_sky_map_z05_bin5}
      \end{figure}

The constant anisotropic model shows a significant improvement over isotropic $w$CDM, with $\Delta(-2\ln\mathcal{L}_{\rm iso}) = 14.75$, consistent with other recent studies~\cite{Verma:2024lex}. 
These gains are not reproduced in permutation null catalogs: across ten shuffled realizations the mean improvement is $\bar{\Delta}=3.6$ with dispersion $\sigma_\Delta=2.9$, and the recovered anisotropy amplitudes/axes are consistent with zero anisotropy.
Although the fit to Pantheon+SH0ES is improved, the predicted CMB quadrupole amplitude significantly exceeds observational constraints, effectively ruling this model out.
The amplitude and redshift evolution of $\delta H/H$ in our 1-bin model (see Fig.~\ref{fig:evolution}) are comparable to the exponentially decaying quadrupole Hubble expansion found by~\cite{Cowell:2022ehf}, suggesting both approaches capture similar underlying anisotropic structure in the data.
For both $w$CDM and constant anisotropic models, $H_0$ is lower than the SH0ES-only preferred value.
DESI prefers slightly \emph{larger} distances than those
implied by SH0ES-anchored SNe.
In these models, the
combined fit can accommodate DESI only by lowering \(H(z)\) broadly at late
times, pushing the model toward a smaller \(H_0\).

Introducing time dependence with five top-hat bins further improves the Pantheon+SH0ES likelihood and permits a higher background rate, $H_0\simeq71\,{\rm km\,s^{-1}\,Mpc^{-1}}$.
As shown in Fig.~\ref{fig:evolution}, the 5-bin fit lifts $w_{\rm eff}$ at very low redshift, boosting $H(z)$ locally and preserving a large $H_0$, while making $w_{\rm eff}$ more negative over $0.2\!\lesssim\!z\!\lesssim\!3.0$ to slow the expansion there and increase the BAO distances $D_M$ and $D_H$.
The total improvement reaches $\Delta(-2\ln\mathcal{L}_{\rm iso})=26.6$ over isotropic $w$CDM.
This improvement arises from two sources: capturing the directional structure in the Pantheon+ data, and alleviating the tension between the SH0ES-anchored distance scale and DESI BAO through redshift-dependent anisotropic stress.
Figure~\ref{fig:anisotropy_sky_map_z05_bin5} shows the corresponding sky pattern at $z\simeq0.5$. Despite the better fit to low-$z$ distances, this freedom also overproduces the ISW quadrupole relative to the Planck constraint, ruling the model out.
The 5-bin models exhibit larger $\delta H/H$ values than reported in~\cite{Cowell:2022ehf}, reflecting the additional freedom that allows the primary expansion directions to reverse sign across redshift bins.

We implement the nonlinear SVD continuation construction for the $N=5$ bin model using $S=10$ continuation steps in the parameter $\lambda$. Because the local null space is estimated numerically (finite-difference derivatives and an SVD at each step) and the Bianchi response is mildly nonlinear, the resulting basis enforces the integrated-shear constraint only approximately. We therefore evaluate the CMB quadrupole for each MCMC sample and impose the quadrupole bound explicitly, retaining only samples that satisfy it (about $10\%$ of draws in this analysis) via rejection sampling. 

The 5-bin SVD model achieves a significant improvement of $\Delta(-2\ln\mathcal{L}_{\rm iso}) = 15.4$ over isotropic $w$CDM while satisfying the CMB quadrupole constraint by construction, with $\log_{10}(C_2^{\rm aniso}) \approx -10$.
The inferred $H_0 = 69.9^{+1.4}_{-0.9}\,{\rm km\,s^{-1}\,Mpc^{-1}}$ lies between the $w$CDM value and the unconstrained 5-bin result, partially alleviating the tension with local distance-ladder measurements.
The underlying equation of state $w_{\rm de} = -0.97^{+0.05}_{-0.06}$ is consistent with a cosmological constant, while the time-varying anisotropic stress produces an effective equation of state that mimics dynamical dark energy.
As shown in Fig.~\ref{fig:evolution}, the constrained model exhibits sign-changing shear that integrates to zero along the line of sight, allowing non-trivial late-time anisotropy without violating the ISW bound.

To quantify how finely the bin amplitudes must be adjusted to satisfy the CMB bound, we performed a local sensitivity test about the best-fit configuration by holding the realized $\Delta w$\,-bin values fixed and perturbing them directly, $c_k\!\to\!c_k\pm\delta$, without recomputing the constrained basis. 
We find that per-bin perturbations at the level of $\delta\sim10^{-4}$ can already push $C_2$ above the CMB limit in some bins (sign-dependent), and coherent joint perturbations of $\pm10^{-4}$ across all bins violate the bound by an order of magnitude. 
Larger steps (e.g. $\delta\gtrsim5\times10^{-4}$) rapidly drive $C_2$ to $\mathcal{O}(10^{-8})$--$10^{-6}$. 
This indicates that maintaining compatibility with the quadrupole bound requires tuning of $\Delta w \sim 10^{-4}$, similar to the result of~\cite{Appleby:2009za}.

\section{Conclusions}

We developed a practical framework for late-time anisotropic dark energy in Bianchi~I that (i) parameterizes time-dependent stress with top–hat bins and (ii) \emph{guarantees} compliance with the CMB temperature quadrupole via an SVD–continuation null–space construction. Applied to Pantheon+SH0ES and DESI BAO, anisotropy improves the fit relative to isotropic $w$CDM. Constant anisotropy is excluded by the ISW quadrupole, but a five–bin constrained model delivers a larger improvement while obeying the bound. Physically, the best fits raise $H(z)$ only at $z\!\lesssim\!0.2$ and drive a more negative effective $w$ at $0.2\!\lesssim\!z\!\lesssim\!3$, mimicking popular dynamical--$w$ solutions.

An important caveat concerns CMB polarization. In our construction we cancel the \emph{temperature} quadrupole by enforcing a vanishing line-of-sight integral of the shear to today. By contrast, large-angle polarization is sourced wherever Thomson scattering occurs in the presence of a \emph{local} quadrupole. In the homogeneous, axisymmetric limit relevant here, that local quadrupole follows the \emph{partial} ISW integral,
\begin{equation}
  Q_T(\eta)\;\equiv\;-\!\int_{\eta_*}^{\eta} s(\eta')\,d\eta'\,.
\end{equation}
Although a time-dependent shear that changes sign can drive cancellations in the temperature integral, the relevant condition for polarization is weighted by the visibility function $g(\eta)=\dot{\tau}e^{-\tau}$, where $\tau$ is the optical depth:
\begin{equation}
  Q_P \;\equiv\; \int_{\eta_*}^{\eta_0} d\eta\, g(\eta)\,Q_T(\eta)\;\approx\;0\,.
\end{equation}
This is a stricter condition, physically distinct from nulling the unweighted temperature integral. In axisymmetric Bianchi~I one does not expect appreciable large-angle \(B\) modes, making \(EE/TE\) the primary observables; a full treatment could augment our constrained-basis scheme to additionally suppress \(Q_P\).

Looking ahead, incorporating polarization constraints and a directional BAO analysis would provide more stringent tests of late-time anisotropy. The effective equation of state $w_{\rm eff}(a)$ generated by anisotropic stress is more constrained than general $w(a)$ parameterizations, limiting the scope for fit improvement. Nonetheless, anisotropic dark energy captures directional structure in the SNe data that isotropic models cannot, with the caveat that CMB compliance requires fine-tuning $\Delta w$ at the $\sim 10^{-4}$ level.

\section*{Acknowledgements}

We would like to thank Ed Copeland, Steven Cotterill and Bryce Cyr for helpful discussions.

\bibliographystyle{apsrev4-1}
\bibliography{refs.bib}

@article{Copeland2006,
    author = "Copeland, Edmund J. and Sami, M. and Tsujikawa, Shinji",
    title = "{Dynamics of dark energy}",
    eprint = "hep-th/0603057",
    archivePrefix = "arXiv",
    doi = "10.1142/S021827180600942X",
    journal = "Int. J. Mod. Phys. D",
    volume = "15",
    pages = "1753--1936",
    year = "2006"
}

@article{DESI:2025fii,
    author = "Lodha, K. and others",
    collaboration = "DESI",
    title = "{Extended Dark Energy analysis using DESI DR2 BAO measurements}",
    eprint = "2503.14743",
    archivePrefix = "arXiv",
    primaryClass = "astro-ph.CO",
    reportNumber = "FERMILAB-PUB-25-0164-PPD",
    month = "3",
    year = "2025"
}

@article{DESI:2025zgx,
    author = "Abdul Karim, M. and others",
    collaboration = "DESI",
    title = "{DESI DR2 Results II: Measurements of Baryon Acoustic Oscillations and Cosmological Constraints}",
    eprint = "2503.14738",
    archivePrefix = "arXiv",
    primaryClass = "astro-ph.CO",
    reportNumber = "FERMILAB-PUB-25-0169-PPD",
    month = "3",
    year = "2025"
}

@article{Planck:2015fie,
    author = "Ade, P. A. R. and others",
    collaboration = "Planck",
    title = "{Planck 2015 results. XIII. Cosmological parameters}",
    eprint = "1502.01589",
    archivePrefix = "arXiv",
    primaryClass = "astro-ph.CO",
    doi = "10.1051/0004-6361/201525830",
    journal = "Astron. Astrophys.",
    volume = "594",
    pages = "A13",
    year = "2016"
}

@article{Planck:2018vyg,
    author = "Aghanim, N. and others",
    collaboration = "Planck",
    title = "{Planck 2018 results. VI. Cosmological parameters}",
    eprint = "1807.06209",
    archivePrefix = "arXiv",
    primaryClass = "astro-ph.CO",
    doi = "10.1051/0004-6361/201833910",
    journal = "Astron. Astrophys.",
    volume = "641",
    pages = "A6",
    year = "2020",
    note = "[Erratum: Astron.Astrophys. 652, C4 (2021)]"
}

@article{SupernovaSearchTeam:1998fmf,
    author = "Riess, Adam G. and others",
    collaboration = "Supernova Search Team",
    title = "{Observational evidence from supernovae for an accelerating universe and a cosmological constant}",
    eprint = "astro-ph/9805201",
    archivePrefix = "arXiv",
    doi = "10.1086/300499",
    journal = "Astron. J.",
    volume = "116",
    pages = "1009--1038",
    year = "1998"
}

@article{SupernovaCosmologyProject:1998vns,
    author = "Perlmutter, S. and others",
    collaboration = "Supernova Cosmology Project",
    title = "{Measurements of $\Omega$ and $\Lambda$ from 42 High Redshift Supernovae}",
    eprint = "astro-ph/9812133",
    archivePrefix = "arXiv",
    reportNumber = "LBNL-41801, LBL-41801",
    doi = "10.1086/307221",
    journal = "Astrophys. J.",
    volume = "517",
    pages = "565--586",
    year = "1999"
}

@article{Appleby:2009za,
    author = "Appleby, Stephen and Battye, Richard and Moss, Adam",
    title = "{Constraints on the anisotropy of dark energy}",
    eprint = "0912.0397",
    archivePrefix = "arXiv",
    primaryClass = "astro-ph.CO",
    doi = "10.1103/PhysRevD.81.081301",
    journal = "Phys. Rev. D",
    volume = "81",
    pages = "081301",
    year = "2010"
}

@article{Battye:2009ze,
    author = "Battye, Richard and Moss, Adam",
    title = "{Anisotropic dark energy and CMB anomalies}",
    eprint = "0905.3403",
    archivePrefix = "arXiv",
    primaryClass = "astro-ph.CO",
    doi = "10.1103/PhysRevD.80.023531",
    journal = "Phys. Rev. D",
    volume = "80",
    pages = "023531",
    year = "2009"
}

@article{Battye:2006mb,
    author = "Battye, Richard A. and Moss, Adam",
    title = "{Anisotropic perturbations due to dark energy}",
    eprint = "astro-ph/0602377",
    archivePrefix = "arXiv",
    doi = "10.1103/PhysRevD.74.041301",
    journal = "Phys. Rev. D",
    volume = "74",
    pages = "041301",
    year = "2006"
}

@article{Pereira:2007yy,
    author = "Pereira, Thiago S. and Pitrou, Cyril and Uzan, Jean-Philippe",
    title = "{Theory of cosmological perturbations in an anisotropic universe}",
    eprint = "0707.0736",
    archivePrefix = "arXiv",
    primaryClass = "astro-ph",
    doi = "10.1088/1475-7516/2007/09/006",
    journal = "JCAP",
    volume = "09",
    pages = "006",
    year = "2007"
}

@article{Verma:2024lex,
    author = "Verma, Anshul and Aluri, Pavan K. and Mota, David F.",
    title = "{Anisotropic universe with anisotropic dark energy}",
    eprint = "2408.08740",
    archivePrefix = "arXiv",
    primaryClass = "astro-ph.CO",
    doi = "10.1103/PhysRevD.111.083508",
    journal = "Phys. Rev. D",
    volume = "111",
    number = "8",
    pages = "083508",
    year = "2025"
}

@article{Scolnic:2021amr,
    author = "Scolnic, Dan and others",
    title = "{The Pantheon+ Analysis: The Full Data Set and Light-curve Release}",
    eprint = "2112.03863",
    archivePrefix = "arXiv",
    primaryClass = "astro-ph.CO",
    doi = "10.3847/1538-4357/ac8b7a",
    journal = "Astrophys. J.",
    volume = "938",
    number = "2",
    pages = "113",
    year = "2022"
}

@article{Foreman-Mackey:2012ig,
    author = "Foreman-Mackey, Daniel and Hogg, David W. and Lang, Dustin and Goodman, Jonathan",
    title = "{emcee: The MCMC Hammer}",
    eprint = "1202.3665",
    archivePrefix = "arXiv",
    primaryClass = "astro-ph.IM",
    doi = "10.1086/670067",
    journal = "Publ. Astron. Soc. Pac.",
    volume = "125",
    pages = "306--312",
    year = "2013"
}

@article{Gorski:2004by,
    author = "G{\'o}rski, K. M. and Hivon, E. and Banday, A. J. and Wandelt, B. D. and Hansen, F. K. and Reinecke, M. and Bartelman, M.",
    title = "{HEALPix - A Framework for high resolution discretization, and fast analysis of data distributed on the sphere}",
    eprint = "astro-ph/0409513",
    archivePrefix = "arXiv",
    doi = "10.1086/427976",
    journal = "Astrophys. J.",
    volume = "622",
    pages = "759--771",
    year = "2005"
}

@article{Saadeh:2016sak,
    author = "Saadeh, Daniela and Feeney, Stephen M. and Pontzen, Andrew and Peiris, Hiranya V. and McEwen, Jason D.",
    title = "{How isotropic is the Universe?}",
    eprint = "1605.07178",
    archivePrefix = "arXiv",
    primaryClass = "astro-ph.CO",
    doi = "10.1103/PhysRevLett.117.131302",
    journal = "Phys. Rev. Lett.",
    volume = "117",
    number = "13",
    pages = "131302",
    year = "2016"
}

@article{Appleby:2015gla,
    author = "Appleby, Stephen A. and Shafieloo, Arman and Johnson, Austin",
    title = "{Direct constraints on the dark energy equation of state from supernovae}",
    eprint = "1512.04079",
    archivePrefix = "arXiv",
    primaryClass = "astro-ph.CO",
    doi = "10.3847/2041-8205/827/1/L6",
    journal = "Astrophys. J. Lett.",
    volume = "827",
    number = "1",
    pages = "L6",
    year = "2016"
}

@article{Colin:2019opn,
    author = "Colin, Jacques and Mohayaee, Roya and Rameez, Mohamed and Sarkar, Subir",
    title = "{Evidence for anisotropy of cosmic acceleration}",
    eprint = "1808.04597",
    archivePrefix = "arXiv",
    primaryClass = "astro-ph.CO",
    doi = "10.1051/0004-6361/201936373",
    journal = "Astron. Astrophys.",
    volume = "631",
    pages = "L13",
    year = "2019"
}

@article{Planck:2019nip,
    author = "Aghanim, N. and others",
    collaboration = "Planck",
    title = "{Planck 2018 results. V. CMB power spectra and likelihoods}",
    eprint = "1907.12875",
    archivePrefix = "arXiv",
    primaryClass = "astro-ph.CO",
    doi = "10.1051/0004-6361/201936386",
    journal = "Astron. Astrophys.",
    volume = "641",
    pages = "A5",
    year = "2020"
}

@article{Motoa-Manzano:2020mwe,
    author = "Motoa-Manzano, J. and Bayron Orjuela-Quintana, J. and Pereira, Thiago S. and Valenzuela-Toledo, C{\'e}sar A.",
    title = "{Anisotropic solid dark energy}",
    eprint = "2012.09946",
    archivePrefix = "arXiv",
    primaryClass = "gr-qc",
    doi = "10.1016/j.dark.2021.100806",
    journal = "Phys. Dark Univ.",
    volume = "32",
    pages = "100806",
    year = "2021"
}

@article{Koivisto:2008xf,
    author = "Koivisto, Tomi and Mota, David F.",
    title = "{Vector Field Models of Inflation and Dark Energy}",
    eprint = "0805.4229",
    archivePrefix = "arXiv",
    primaryClass = "astro-ph",
    doi = "10.1088/1475-7516/2008/08/021",
    journal = "JCAP",
    volume = "08",
    pages = "021",
    year = "2008"
}

@article{Orjuela-Quintana:2022jrg,
    author = "Orjuela-Quintana, J. Bayron and Palacios-C{\'o}rdoba, Jose L. and Valenzuela-Toledo, C{\'e}sar A.",
    title = "{Late-time anisotropy sourced by a 2-form field non-minimally coupled to cold dark matter}",
    eprint = "2202.07546",
    archivePrefix = "arXiv",
    primaryClass = "gr-qc",
    doi = "10.1016/j.dark.2024.101575",
    journal = "Phys. Dark Univ.",
    volume = "46",
    pages = "101575",
    year = "2024"
}

@article{Alhulaimi:2013sha,
    author = "Alhulaimi, B. and Coley, A. and Sandin, P.",
    title = "{Anisotropic Einstein-aether cosmological models}",
    doi = "10.1063/1.4802246",
    journal = "J. Math. Phys.",
    volume = "54",
    pages = "042503",
    year = "2013"
}

@article{Rubin:2023jdq,
    author = "Rubin, David and others",
    title = "{Union Through UNITY: Cosmology with 2,000 SNe Using a Unified Bayesian Framework}",
    eprint = "2311.12098",
    archivePrefix = "arXiv",
    primaryClass = "astro-ph.CO",
    doi = "10.3847/1538-4357/adc0a5",
    journal = "Astrophys. J.",
    volume = "986",
    number = "2",
    pages = "231",
    year = "2025"
}

@article{DES:2024jxu,
    author = "Abbott, T. M. C. and others",
    collaboration = "DES",
    title = "{The Dark Energy Survey: Cosmology Results with {\ensuremath{\sim}}1500 New High-redshift Type Ia Supernovae Using the Full 5 yr Data Set}",
    eprint = "2401.02929",
    archivePrefix = "arXiv",
    primaryClass = "astro-ph.CO",
    reportNumber = "FERMILAB-PUB-23-0821-PPD, DES-2023-805",
    doi = "10.3847/2041-8213/ad6f9f",
    journal = "Astrophys. J. Lett.",
    volume = "973",
    number = "1",
    pages = "L14",
    year = "2024"
}

@article{Ellis:1968vb,
    author = "Ellis, G. F. R. and MacCallum, Malcolm A. H.",
    title = "{A Class of homogeneous cosmological models}",
    doi = "10.1007/BF01645908",
    journal = "Commun. Math. Phys.",
    volume = "12",
    pages = "108--141",
    year = "1969"
}

@article{Sah:2024csa,
    author = "Sah, Animesh and Rameez, Mohamed and Sarkar, Subir and Tsagas, Christos G.",
    title = "{Anisotropy in Pantheon+ supernovae}",
    eprint = "2411.10838",
    archivePrefix = "arXiv",
    primaryClass = "astro-ph.CO",
    doi = "10.1140/epjc/s10052-025-14222-w",
    journal = "Eur. Phys. J. C",
    volume = "85",
    number = "5",
    pages = "596",
    year = "2025"
}

@article{Cowell:2022ehf,
    author = "Cowell, Jessica A. and Dhawan, Suhail and Macpherson, Hayley J.",
    title = "{Potential signature of a quadrupolar hubble expansion in Pantheon+supernovae}",
    eprint = "2212.13569",
    archivePrefix = "arXiv",
    primaryClass = "astro-ph.CO",
    doi = "10.1093/mnras/stad2788",
    journal = "Mon. Not. Roy. Astron. Soc.",
    volume = "526",
    number = "1",
    pages = "1482--1494",
    year = "2023"
}

@article{Sorrenti:2022zat,
    author = "Sorrenti, Francesco and Durrer, Ruth and Kunz, Martin",
    title = "{The dipole of the Pantheon+SH0ES~data}",
    eprint = "2212.10328",
    archivePrefix = "arXiv",
    primaryClass = "astro-ph.CO",
    doi = "10.1088/1475-7516/2023/11/054",
    journal = "JCAP",
    volume = "11",
    pages = "054",
    year = "2023"
}

\vskip 0.5cm

\appendix

\section{Constrained Basis Construction}
\label{sec:cons}

We construct basis functions that ensure the integrated shear is close to zero, i.e.\ $Q_T \approx 0$, so that the CMB quadrupole constraint is satisfied while allowing late-time anisotropy. Using $\mathrm{d}\eta/\mathrm{d}a=(a\mathcal{H})^{-1}$, the CMB constraint is
\begin{equation}
  Q_T \;=\; -\!\int_{a_*}^{a_0} \frac{s(a)}{a\,\mathcal{H}(a)}\,\mathrm{d}a \;\approx\; 0,
  \qquad a_0=1.
  \label{eq:Q_constraint}
\end{equation}
The shear has the integral solution
\begin{equation}
  s(a) \;=\; 8\pi G\,a^{-2}\!\int_{a_*}^{a}
  \frac{a'^{3}\,\rho_{\rm de}(a')}{\mathcal{H}(a')}\;
  \Delta w(a')\,\mathrm{d}a'\,,
  \label{eq:shear_integral}
\end{equation}
where $s(a_*)=0$. For each bin $B_k$ in (\ref{eq:dw_bins}), we define
\begin{equation}
  Q_k \;\equiv\; -\!\int_{a_*}^{a_0} \frac{s_k(a)}{a\,\mathcal{H}(a)}\,\mathrm{d}a\,,
  \label{eq:Qk_def}
\end{equation}
where $s_k$ is obtained by solving the Bianchi system with unit amplitude in bin $k$ only. For sufficiently small $\Delta w$, the response is linear and
\begin{equation}
  Q_T \;\approx\; \sum_{k=0}^{N-1} c_k\,Q_k\,.
\end{equation}
Enforcing \eqref{eq:Q_constraint} at linear order gives the single linear constraint
\begin{equation}
  \sum_{k=0}^{N-1} c_k\,Q_k \;=\; 0\,.
  \label{eq:Q_row_constraint}
\end{equation}
Defining $\mathbf{Q}=[Q_0,\dots,Q_{N-1}]\in\mathbb{R}^{1\times N}$ and $\mathbf{c}=[c_0,\dots,c_{N-1}]^{\!\top}\in\mathbb{R}^{N}$, we have $\mathbf{Q}\,\mathbf{c}=0$, which means that $\mathbf{c}$ lies in the null space of the $1\times N$ row $\mathbf{Q}$.

To construct this null space, we compute the singular value decomposition
\begin{equation}
  \mathbf{Q} \;=\; \mathbf{U}\,\mathbf{S}\,\mathbf{V}^{\!\top},
  \qquad \mathbf{U}\in\mathbb{R}^{1\times 1},\ \mathbf{S}\in\mathbb{R}^{1\times 1},\ \mathbf{V}\in\mathbb{R}^{N\times N}.
\end{equation}
The null space is then spanned by the $N-1$ right singular vectors corresponding to zero singular values, which correspond to columns $2,\ldots,N$ of $\mathbf{V}$.
We collect these columns as $V_{\rm null}\in\mathbb{R}^{N\times(N-1)}$, so that any admissible coefficient vector can be written as
\begin{equation}
  \mathbf{c} \;=\; \sum_{j=1}^{N-1} \alpha_j\,\mathbf{v}_j \;=\; V_{\rm null}\,\boldsymbol{\alpha},
  \qquad \boldsymbol{\alpha}=[\alpha_1,\ldots,\alpha_{N-1}]^{\!\top},
  \label{eq:c_expansion_svd}
\end{equation}
and the corresponding basis functions are
\begin{equation}
  \Delta w(a) \;=\; \sum_{j=1}^{N-1} \alpha_j\,\tilde{B}_j(a)\,,
  \qquad
  \tilde{B}_j(a) \;=\; \sum_{k=0}^{N-1} (\mathbf{v}_j)_k\,B_k(a)\,.
  \label{eq:final_dw_parameterization_svd}
\end{equation}

For larger amplitudes, $\Delta w$ modifies $\mathcal{H}(a)$ and $s(a)$ nonlinearly, causing the linear null directions to rotate with amplitude. To handle this, we introduce an amplitude parameter $\lambda\in[0,1]$ and write
\begin{equation}
  \Delta w(a;\lambda,\mathbf{c}) \;=\; \lambda \sum_{k=0}^{N-1} c_k\,B_k(a)
  \;=\; \lambda\,\mathbf{c}^{\!\top}\mathbf{B}(a),
  \label{eq:dw_lambda}
\end{equation}
with $\mathbf{B}(a)\equiv[B_0(a),\ldots,B_{N-1}(a)]^{\!\top}$. We then increase $\lambda$ from $0$ to $1$ in $S$ steps of size $\Delta\lambda=1/S$, recomputing the local null space at each step.

At any finite-amplitude base point $(\lambda,\mathbf{c})$, we approximate the local row response of $Q$ to the bin coefficients using central finite differences:
\begin{equation}
  Q^{\rm loc}_k(\lambda,\mathbf{c})
  \;\approx\;
  \frac{Q\!\left(\lambda,\mathbf{c}+\varepsilon\,\mathbf{e}_k\right)-Q\!\left(\lambda,\mathbf{c}-\varepsilon\,\mathbf{e}_k\right)}{2\varepsilon}\,,
  \label{eq:Qloc_fd}
\end{equation}
where $\mathbf{e}_k$ is the $k$-th Euclidean basis vector. Collecting these entries yields the $1\times N$ row $\mathbf{Q}^{\rm loc}(\lambda,\mathbf{c})=[Q^{\rm loc}_0,\dots,Q^{\rm loc}_{N-1}]$.
We then compute the SVD of $\mathbf{Q}^{\rm loc}$ and extract the $(N-1)$-dimensional right-null basis $V_{\rm null}(\lambda,\mathbf{c})\in\mathbb{R}^{N\times(N-1)}$.

Given a fixed parameter vector $\boldsymbol{\alpha}\in\mathbb{R}^{N-1}$, we form a direction in the local null space as
\[
  \mathbf{d} \;=\; V_{\rm null}(\lambda,\mathbf{c})\,\boldsymbol{\alpha} \;\in\; \mathbb{R}^{N},
\]
and perform the update step
\begin{equation}
  \lambda \;\leftarrow\; \lambda + \Delta\lambda,
  \qquad
  \mathbf{c} \;\leftarrow\; \mathbf{c} + \Delta\lambda\,\mathbf{d}.
  \label{eq:cont_update}
\end{equation}
Starting from the initial conditions $\lambda=0$ and $\mathbf{c}=\mathbf{0}$, and repeating the process described in \eqref{eq:Qloc_fd}–\eqref{eq:cont_update} for $S$ steps, we obtain $\Delta w(a)$ at $\lambda=1$ that follows the evolving local null space of $Q$.

\begin{figure}[t!]
\centering
\includegraphics[width=0.45\textwidth]{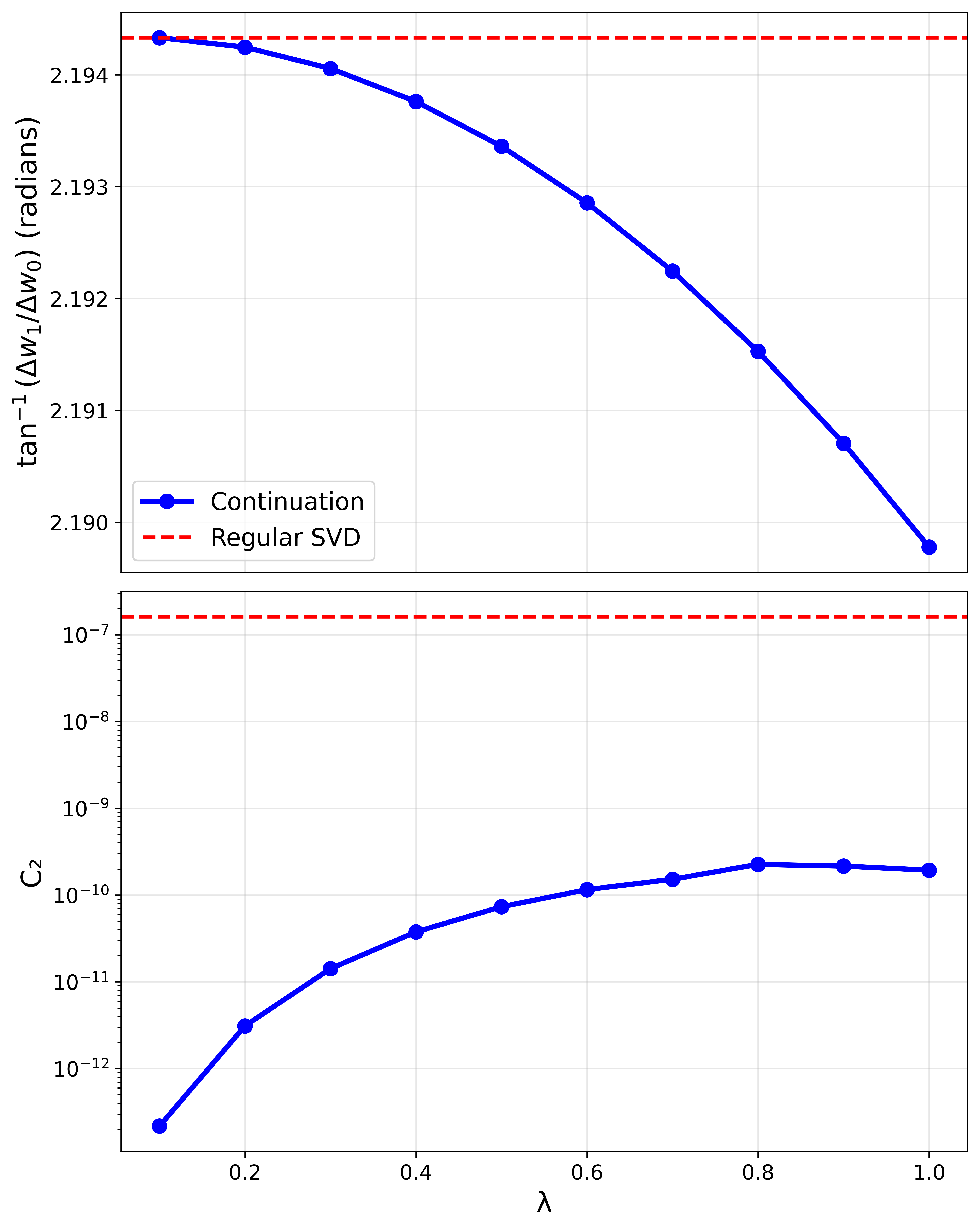}
\caption{Demonstration of the SVD continuation method for a 2-bin parameterization with $\alpha_1 = 0.3$. \textit{Upper panel:} Evolution of the bin angle $\tan^{-1}(\Delta w_1/\Delta w_0)$ as a function of the continuation parameter $\lambda$. The blue line shows how the direction evolves during continuation, while the red dashed line indicates the fixed direction from the standard SVD method. \textit{Lower panel:} Evolution of $C_2$ during continuation. The continuation method (blue) maintains much better constraint satisfaction than the standard SVD approach (red dashed line).}
\label{fig:svd_continuation}
\end{figure}

Figure~\ref{fig:svd_continuation} demonstrates the nonlinear continuation approach compared to the standard SVD method. We consider a simplified 2-bin parameterization ($N=2$) with one null space dimension and coefficient $\alpha_1 = 0.3$. The upper panel shows the evolution of the angle between the two bins, parameterized as $\tan^{-1}(\Delta w_1/\Delta w_0)$, as a function of $\lambda$. The lower panel shows the evolution of the CMB quadrupole coefficient $C_2$. The continuation method achieves significantly better constraint satisfaction, with $C_2$ values several orders of magnitude smaller than the standard SVD approach.

\section{Derivation of $w_{\rm eff}$} \label{sec:weff}

The effective dark energy density is given by \eqref{eq:rho_de_eff}, and the corresponding effective equation of state by
\begin{equation}
  \rho_{\rm de}^{\rm (eff)\prime}
  = -3\mathcal{H}\bigl(1+w_{\rm eff}\bigr)\rho_{\rm de}^{\rm (eff)}.
\end{equation}
Using the background evolution equations for $\rho_{\rm de}$ and
$\sigma_i^{\;j}$ one finds
\begin{equation}
  \rho_{\rm de}^{\rm (eff)\prime}
  = -3\mathcal{H}\Bigl[(1+w_{\rm de})\,\rho_{\rm de}
                       + 2\rho_\sigma\Bigr],
\end{equation}
so that equating the two expressions for
$\rho_{\rm de}^{\rm (eff)\prime}$ gives
\begin{equation}
  (1+w_{\rm eff})(\rho_{\rm de}+\rho_\sigma)
  = (1+w_{\rm de})\rho_{\rm de} + 2\rho_\sigma.
\end{equation}
Hence
\begin{equation}
  w_{\rm eff}
  = \frac{w_{\rm de}\,\rho_{\rm de} + \rho_\sigma}
         {\rho_{\rm de}+\rho_\sigma},
\end{equation}
which can be rearranged as
\begin{equation}
  w_{\rm eff} - w_{\rm de}
  = \frac{(1-w_{\rm de})\,\rho_\sigma}
         {\rho_{\rm de}+\rho_\sigma}.
\end{equation}
For $\rho_\sigma \ge 0$ and $w_{\rm de}\le 1$ this implies
$w_{\rm eff}\ge w_{\rm de}$, i.e.\ the shear contribution always
increases the effective equation of state relative to the underlying
dark energy.

\section{Null tests} \label{sec:null}

To assess whether the improvement of the anisotropic model over the isotropic baseline could arise from survey geometry or chance alignments, we construct permutation-based null catalogs that preserve the data selection while destroy\-ing only the angular information exploited by the model. 
For each unique survey identifier in Pantheon+SH0ES we randomly permute the sky coordinates among the \emph{cosmological} SNe, leaving all other observables and covariances unchanged (e.g.\ \(m_{B,\mathrm{corr}},x_1,c\), redshifts, host properties, velocity and extinction terms). 
This stratified shuffle preserves each survey’s footprint, depth, calibration, and redshift distribution, thereby generating realizations consistent with the isotropic null given the actual selection function.

We then perform full posterior inference on ten such randomized catalogs, refitting the single-bin constant-anisotropy model (including the sky axis) each time. 
For catalog $j$ we define the likelihood-ratio improvement relative to this baseline as
\[
\Delta_j \equiv -2\!\left[\ln\mathcal{L}_{\rm iso}-\ln\mathcal{L}_{\rm aniso}\right],
\]
where $\mathcal{L}_{\rm iso}$ is the maximum likelihood under isotropic $w$CDM and $\mathcal{L}_{\rm aniso}$ is the maximum likelihood under the anisotropic model for the same catalog. Because the anisotropic model nests the isotropic $w$CDM baseline, the likelihood-ratio statistic is non-negative by construction, and a modest positive gain is expected under the null simply from the additional flexibility.  

Across the ten null runs we find \(\Delta\) consistent with the isotropic baseline, with sample mean \(\bar{\Delta}=3.6\) and standard deviation \(\sigma_\Delta=2.9\). 
The recovered anisotropy amplitude $c_1$ is consistent with zero across the shuffled catalogs.
Angular-uniformity tests of the axis posteriors (prior‑consistent \(u_\phi\!=\!\phi/2\pi\), \(u_\theta\!=\!\theta/(\pi/2)\) with heavy thinning, calibrated to an empirical null from permutation catalogs) find the shuffled catalogs consistent with isotropy, whereas the real data show a strong azimuthal deviation (bin1) and strong azimuthal \emph{and} polar deviations (bin5).
Together these results show no evidence that the constrained basis induces spurious anisotropy when sky directions are randomized.

\end{document}